\documentclass{svproc}

\usepackage{amsmath}
\usepackage{amsfonts}
\usepackage{amssymb}
\usepackage{graphicx}
\usepackage{pythonhighlight}
\usepackage{subfig}
\usepackage{url}

\begin{document}

\mainmatter             	

\title{A note on a prey-predator model\\ with constant-effort harvesting\thanks{This is a preprint 
whose final form is published by Springer Nature Switzerland AG in the book 
'Dynamic Control and Optimization'. Submitted 30/Nov/2021; Accepted 10/Feb/2022.}}

\titlerunning{A note on a prey-predator model with constant-effort harvesting} 

\author{M\'{a}rcia Lemos-Silva \and Delfim F. M. Torres}

\authorrunning{M. Lemos-Silva and D. F. M. Torres} 

\tocauthor{M\'{a}rcia Lemos-Silva and Delfim F. M. Torres}

\institute{Center for Research and Development in Mathematics and Applications (CIDMA),
Department of Mathematics, University of Aveiro, 3810-193 Aveiro, Portugal\\
\email{marcialemos@ua.pt} and \email{delfim@ua.pt}}

\maketitle             


\begin{abstract}
We study a prey-predator model based on the classical Lotka--Volterra system 
with Leslie--Gower and Holling IV schemes and a constant-effort harvesting. 
Our goal is twofold: to present the model proposed by Cheng and Zhang in 2021,
pointing out some inconsistencies; to analyse the number and type 
of equilibrium points of the model. We end by proving the stability 
of the meaningful equilibrium point, according to the distribution 
of the eigenvalues.


\medskip

\end{abstract}


\section{Introduction}

Prey-predator equations describe an ecological system of two linked species 
that depend on each other. One is the prey, which provides food for the other, 
the predator. Under some conditions, both prey and predator populations grow. 
Lotka (1880--1949) studied such equations in his book of 1925 \cite{Lotka:1925}; 
Volterra (1860--1940) investigated them, independently \cite{Volterra:1928}; 
and, for this reason, such prey-predator equations 
are also known as Lotka--Volterra equations. 
Recently, there has been a tremendous amount 
of research done in this area \cite{MR3157463}.

Here we consider a prey-predator model with Leslie--Gower 
and Holling IV schemes with constant-effort harvesting,
proposed by Cheng and Zhang in 2021 \cite{cheng:zhang}.
The Cheng--Zhang model is given by
\begin{equation}
\label{eq:1}
\begin{cases}
\frac{dx}{dt} &= r_1 x \left(1-\frac{x}{K}\right) - \frac{mx}{b+x^2}y - c_1x,\\
\frac{dy}{dt} &= r_2 y \left(1 - \frac{y}{sx}\right) - c_2y,
\end{cases}
\end{equation}
where $x(t)$ and $y(t)$ represent the size at time $t$
of the prey and the predator populations, respectively; 
$K$ denotes the environmental carrying capacity for the prey; 
$m$ is the maximal predation rate; $s$ measures the quality 
of the prey as food for the predator; $b$ denotes the half-saturation constant, 
i.e., it measures the resources availability at which half of the maximum intake 
is reached; $c_1$ and $c_2$ measure the harvesting efforts; and $r_1$ and $r_2$ 
are the intrinsic growth rates of the prey and predators, respectively. 

In the first equation of \eqref{eq:1}, a logistic model $r_1 x\left(1 - \frac{x}{K}\right)$ 
is used to describe the growth of the prey when there are no predators in an environment, 
which is limited by the carrying capacity $K$; the simplified Holling IV response, given 
by the term $\frac{mx}{b+x^2}y$, describes the density of the prey attacked by the predators 
per unit of time; while $c_1 x$ denotes the constant-effort harvesting of the prey. 
In the second equation of \eqref{eq:1}, the Leslie--Gower function $r_2 y\left(1 - \frac{y}{sx}\right)$ 
is used to describe the growth of the predators and $c_2 y$ 
represents the constant-effort harvesting of the predators. 

In the past forty years, Computer Algebra Systems (CAS) have drastically 
changed the everyday practice of mathematics \cite{MR4119620}.
Here we use the free and open-source CAS \textsf{SageMath} \cite{sagemath} 
to give a simple and direct analysis of the prey-predator dynamical system. 
The obtained results show inaccuracies to the conclusions in \cite{cheng:zhang}
that may jeopardize the model. We conclude that \textsf{SageMath} is a wonderful
tool to guarantee reproducible results and to avoid mistakes in the calculations.


\section{An equivalent model}

Let us consider the one-to-one scaling transformations
\begin{equation}
\label{eq:transf}
\overline{t} = r_1t, 
\quad \overline{x} = \frac{x}{K}, 
\quad \overline{y} = \frac{my}{r_1K^2}, 
\end{equation}
and the new quantities
\begin{equation}
\label{eq:new:parameters}
\quad a := \frac{b}{K^2}, 
\quad \delta := \frac{r_2}{r_1}, 
\quad \beta := \frac{r_1K}{sm}, 
\quad h_1 := \frac{c_1}{r_1}, 
\quad h_2 := \frac{c_2}{r_1},
\end{equation}
defined from the parameters of model \eqref{eq:1}.

\begin{remark}
The expressions of $\beta$ and $h_2$ 
given in \cite{cheng:zhang} have a typo.
\end{remark}

One has from \eqref{eq:transf} that 
\begin{gather*}
d\overline{t} = r_1 \ dt \ \Leftrightarrow \ dt \ = \frac{1}{r_1} \ d\overline{t},\\
d\overline{x} = \frac{1}{K} \ dx \  \Leftrightarrow  \ dx \ = K \ d\overline{x},\\
d\overline{y} = \frac{m}{r_1 K^2} \ dy \ \Leftrightarrow \ dy \ = \frac{r_1K^2}{m} \ d\overline{y}.
\end{gather*}
Therefore,
\begin{equation*}
\frac{dx}{dt} = \frac{K\ d\overline{x}}{\frac{1}{r_1}d\overline{t}} 
= Kr_1 \frac{d\overline{x}}{d\overline{t}}
\end{equation*}
and
\begin{equation*}
\frac{dy}{dt} = \frac{\frac{r_1K^2}{m}d\overline{y}}{\frac{1}{r_1}d\overline{t}} 
= \frac{(r_1K)^2}{m}\frac{d\overline{y}}{d\overline{t}}.
\end{equation*}
This means that we can rewrite the first equation 
$\frac{dx}{dt} = r_1 x \left(1-\frac{x}{K}\right) - \frac{mx}{b+x^2}y - c_1x$
of system \eqref{eq:1} in the new variables as follows:
$$
Kr_1\frac{d\overline{x}}{d\overline{t}} 
= r_1K\overline{x} \left(1- \overline{x}\right) 
- \frac{mK\overline{x}}{b + (K\overline{x})^2}
\times\frac{r_1K^2\overline{y}}{m} - c_1K\overline{x},
$$
which is equivalent to
\begin{equation*}
\begin{split}
\frac{d\overline{x}}{d\overline{t}} 
&= \overline{x}\left(1-\overline{x}\right) 
- \frac{m\overline{x}}{b + (K\overline{x})^2} 
\times \frac{K^2\overline{y}}{m} - \frac{c_1}{r_1}\overline{x}\\
&= \overline{x}(1-\overline{x}) 
- \frac{\overline{x}K^2\overline{y}}{b + K^2\overline{x}^2} - h_1\overline{x}\\
&= \overline{x}(1-\overline{x}) - \frac{\overline{x}}{\frac{b}{K^2} 
+ \overline{x}^2}\overline{y} - h_1\overline{x}\\
&= \overline{x}(1-\overline{x}) 
- \frac{\overline{x}}{a + \overline{x}^2}\overline{y} - h_1\overline{x}.
\end{split}	
\end{equation*}
Similarly, the second equation 
$\frac{dy}{dt} = r_2 y \left(1 - \frac{y}{sx}\right) - c_2y$
of system \eqref{eq:1} is given in the new variables by
$$
\frac{(r_1K)^2}{m}\frac{d\overline{y}}{d\overline{t}} 
= \frac{r_2r_1K^2\overline{y}}{m}
\left(1 - \frac{\frac{r_1K^2\overline{y}}{m}}{s K\overline{x}}\right) 
- \frac{c_2r_1K^2\overline{y}}{m},
$$
which is equivalent to
\begin{equation*}
\begin{split}
\frac{d\overline{y}}{d\overline{t}} 
&= \frac{r_2}{r_1}\overline{y}\left(1 - \frac{r_1K\overline{y}}{ms\overline{x}}\right) 
- \frac{c_2}{r_1}\overline{y}\\
&= \delta\overline{y}\left(1 - \beta \frac{\overline{y}}{\overline{x}}\right) 
- h_2\overline{y}.
\end{split}	
\end{equation*}
We conclude that system \eqref{eq:1} is given, in the new variables, as
\begin{equation}
\label{eq:2}
\begin{cases}
\frac{d\overline{x}}{d\overline{t}} 
= \overline{x}(1-\overline{x}) 
- \frac{\overline{x}}{a + \overline{x}^2}\overline{y} - h_1\overline{x},\\
\frac{d\overline{y}}{d\overline{t}} 
= \delta\overline{y}\left(1 - \beta 
\frac{\overline{y}}{\overline{x}}\right) - h_2\overline{y},
\end{cases}
\end{equation}
where $a$, $\delta$, $\beta$, $h_1$ and $h_2$ 
are the positive rescaled parameters 
given by \eqref{eq:new:parameters}. Systems \eqref{eq:1}
and \eqref{eq:2} are equivalent and we proceed
by analysing \eqref{eq:2}.

\begin{remark}
The second equation of system \eqref{eq:2}
is wrongly written in \cite{cheng:zhang} as
$$
\frac{d\overline{y}}{d\overline{t}} 
= \overline{y}\left(\delta - \beta 
\frac{\overline{y}}{\overline{x}}\right) - h_2\overline{y}
$$
(cf. system (1.2) of \cite{cheng:zhang}).
\end{remark}


\section{Equilibria and stability}

In this section, we analyse the number and type of equilibria 
for system \eqref{eq:2} and prove the stability of the meaningful
equilibrium point. We make use of the free open-source 
mathematics software system \textsf{SageMath} \cite{sagemath}.  

We start by calculating the equilibrium points using the script
\begin{python}
	var('x,y,a,h1,delta,h2,beta')
	eq1 = x*(1-x) - (x/(a+x^2))*y - h1*x 
	eq2 = delta*y*(1-beta*(y/x))-h2*y 
	pretty_print(solve((eq1,eq2),(x,y)))
\end{python}
\indent from which we obtain four possible equilibria: 
$(0,0)$, $(-h_1+1,0)$, $(-\sqrt{-a},0)$, $(\sqrt{-a},0)$. 
As we are working with a prey-predator model, from the perspective of ecology, 
we are only interested in the pairs $(\overline{x},\overline{y})$ such that 
$\overline{x} \geq 0$ and $\overline{y}\geq 0$. As $a$ is a positive parameter and 
$(0,0)$ means the extinction of both species, it is simple to understand that 
the only feasible equilibrium point we are interested in studying is $E = (1-h_1,0)$
with $0\leq h_1<1$. Indeed, if both populations are at 0 ($h_1 = 1$), 
then they will continue to be so indefinitely.
Unfortunately, and in contrast with the classical Lotka--Volterra model, here we 
do not have a fixed point at which both populations sustain their non-zero numbers.

\begin{remark}
In \cite{cheng:zhang} the authors claim
the existence of a positive equilibrium $(x^{*},y^{*})$
of the system with $x^{*}>0$ and $y^{*}>0$. Unfortunately 
there is no such equilibrium, which means that
the model \eqref{eq:1} proposed by Cheng and Zhang in 2021
is not realistic.
\end{remark}

To determine the type of equilibrium, we first calculate 
the Jacobian matrix $J$ evaluated at the point $(-h_1 + 1, 0)$. 
We use the following \textsf{SageMath} script:
\begin{python}
	j = jacobian((eq1,eq2),(x,y))
	jac = j.substitute(x = -h1+1, y = 0)
	jac 
\end{python}
\noindent obtaining
$$
\left(\begin{array}{cc}
h_{1} - 1 & \frac{h_{1} - 1}{{\left(h_{1} - 1\right)}^{2} + a} \\
0 & \delta - h_{2}
\end{array}\right).
$$
The corresponding eigenvalues are obtained using the \textsf{SageMath} command 
\begin{verbatim}
  jac.eigenvalues() 
\end{verbatim}  
By doing so, we get the following two eigenvalues: 
$\delta - h_2$ and $h_1 - 1$. We have just proved the following result.

\begin{theorem}
\label{thm01}
If $h_1 \in \, ]0, 1[$ and $\delta \in \, ]0, h_2[$, then
the equilibrium point $(1-h_1,0)$ is a sink.
\end{theorem}


\section{Numerical simulations}

Now we use \textsf{SageMath} to plot some solutions of the
prey-predator model and illustrate the fact that for different 
initial values, if one chooses the parameters according 
with Theorem~\ref{thm01}, then the solutions of the model converge to the 
equilibrium point $(1-h_1, 0)$. For this purpose, we start by importing 
some libraries for the numerical integration of the non-linear system
\eqref{eq:2} and the visualization of its solutions:
\begin{python}
	import numpy as np 
	from scipy import integrate
	from scipy.integrate import odeint
	import matplotlib.pyplot as plt
\end{python}
\noindent Then we define model \eqref{eq:2} with
\begin{equation*}
\quad a := 0.2, 
\quad \delta := 0.3, 
\quad \beta := 0.8, 
\quad h_1 := 0.4, 
\quad h_2 := 0.6,
\end{equation*}
as follows:
\begin{python}
	def model (z,t):
		x = z[0]
		y = z[1]
		dxdt = x*(1-x) - (x/(0.2+x^2))*y - 0.4*x
		dydt = y*0.3*(1 - 0.8*(y/x)) - 0.6*y
		dzdt = [dxdt,dydt]
		return dzdt	
\end{python}
Next, we compute the solutions of the system 
for initial conditions 
$(\overline{x}(0),\overline{y}(0))$ 
given by $(3,1)$, $(2,4)$ and $(4,3)$:
\begin{python}
	z0 = [3,1]
	z1 = [2,4]
	z2 = [4,3]
	n = 5
	t = np.linspace(0,50,n)
	x = np.empty_like(t)
	xx = np.empty_like(t)
	xxx = np.empty_like(t)
	y = np.empty_like(t)
	yy = np.empty_like(t)
	yyy = np.empty_like(t)
	x[0] = z0[0]
	xx[0] = z1[0]
	xxx[0] = z2[0]
	y[0] = z0[1]
	yy[0] = z1[1]
	yyy[0] = z2[1]
	
	for i in range(1,n):
		tspan = [t[i-1],t[i]]
		z = odeint(model,z0,tspan)
		zz = odeint(model,z1,tspan)
		zzz = odeint(model,z2,tspan)
		x[i] = z[1][0]
		xx[i] = zz[1][0]
		xxx[i] = zzz[1][0]
		y[i] = z[1][1]
		yy[i] = z[1][1]
		yyy[i] = z[1][1]
		z0 = z[1]
		z1 = zz[1]
		z2 = zzz[1]
\end{python}
\noindent Finally, we plot the solutions with

\begin{figure}[ht!]
\centering
\subfloat[Prey]{\includegraphics[scale=0.6]{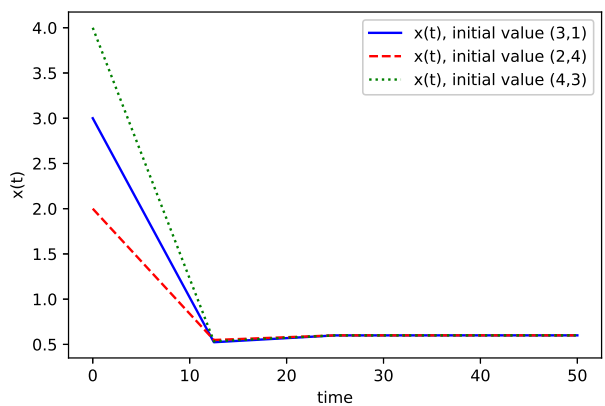}}\\
\subfloat[Predator]{\includegraphics[scale=0.6]{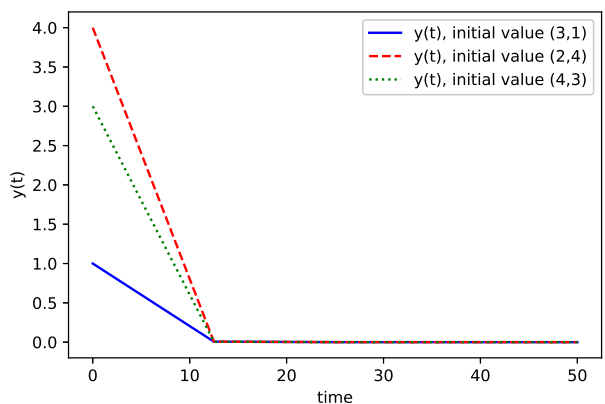}}
\caption{Solutions of system \eqref{eq:2} with 
$a = 0.2$, $h_1 = 0.4$, $\delta = 0.3$, $\beta = 0.8$, $h_2 = 0.6$
and different initial conditions $(\overline{x}(0),\overline{y}(0))$: 
$(3,1)$, $(2,4)$ and $(4,3)$.}
\label{Fig.1}
\end{figure}

\begin{python}
	plt.figure(figsize = (10,4))
	plt.subplot(1,2,1)
	plt.plot(t,x,'b-',label='x(t), initial value (3,1)', alpha=0.5)
	plt.plot(t,xx,'r--',label='x(t), initial value (2,4)', alpha = 0.5)
	plt.plot(t,xxx, 'g:', label='x(t), initial value (4,3)', alpha = 1)
	plt.ylabel('x(t)')
	plt.xlabel('time')
	plt.legend(loc='best')
	
	plt.subplot(1,2,2)
	plt.plot(t,y,'b-',label='y(t), initial value (3,1)', alpha = 0.5)
	plt.plot(t,yy,'r--',label='y(t), initial value (2,4)', alpha = 0.5)
	plt.plot(t,yyy, 'g:', label='y(t), initial value (4,3)', alpha = 1)
	plt.ylabel('y(t)')
	plt.xlabel('time')
	plt.legend(loc='best')
\end{python}
obtaining Figure~\ref{Fig.1}.

\begin{figure}[ht!]
\centering
\includegraphics[scale = 0.6]{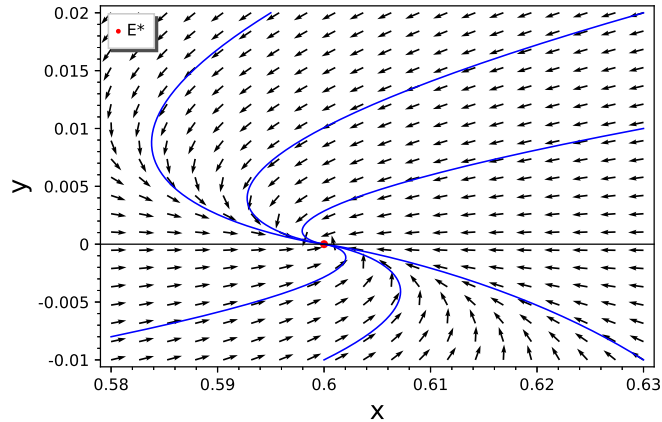} 
\caption{Phase portrait for system \eqref{eq:2} with $a = 0.2$, 
$h_1 = 0.4$, $\delta = 0.3$, $\beta = 0.8$ and $h_2 = 0.6$.}
\label{Fig.2}
\end{figure}

We can also illustrate the stability of the meaningful equilibrium point 
of the prey-predator system by plotting the phase portrait of \eqref{eq:2}. 
For that, we use random initial conditions through the following script:

\begin{python}
	a,h1,delta,beta,h2 = 0.2,0.4,0.1,0.8,0.6
	def dX_dt(X, t=0): 
		return [X[0]*(1-X[0]) - (X[0]/(a+X[0]^2))*X[1] - h1*X[0], X[1]*(delta-beta*(X[1]/X[0])) - h2*X[1]]
	
	def g(x,y):
		v = vector(dX_dt([x, y])) 
		return v/v.norm() 
	
	var('x,y')
	v = plot_vector_field(g(x,y), (x,0.58,0.63), (y,-0.01,0.02), axes_labels = ("x", "y"))
	
	t = srange(0, 15, .01)
	X = integrate.odeint(dX_dt, [0.63,0.01], t) 
	X1 = integrate.odeint(dX_dt, [0.63,0.02], t)
	X2 = integrate.odeint(dX_dt, [0.595,0.02], t)
	X3 = integrate.odeint(dX_dt, [0.58, -0.008], t)
	X4 = integrate.odeint(dX_dt, [0.6, -0.01], t)
	X5 = integrate.odeint(dX_dt, [0.63, -0.01], t)
	q = line(X) + line(X1) + line(X2) + line(X3) + line(X4) + line(X5)
	
	p = points((0.6,0), color='red', legend_label='E*', size = 25)
	
	show(v+p+q)
\end{python}
\noindent The obtained phase portrait is displayed in Figure~\ref{Fig.2}.


\section{Conclusion}
\label{sec:conc}

In this paper we considered a prey-predator model 
with constant harvesting effort previously studied 
in \cite{cheng:zhang}. We remarked several typos 
and inconsistencies in \cite{cheng:zhang}, showing 
detailed computations for each of them. We trust
that the analysis of the stability of the equilibrium points 
may be helpful to new researchers in the field.  
As future work, we plan to modify the proposed model 
in order to make possible the co-existence equilibrium point. 


\section*{Acknowledgments}

The authors are grateful to the financial support 
of The Center for Research and Development 
in Mathematics and Applications (CIDMA) through the
Portuguese Foundation for Science and Technology 
(FCT -- Funda\c{c}\~{a}o para a Ci\^{e}ncia e a Tecnologia),
projects UIDP/04106/2020 (Lemos-Silva) and UIDB/04106/2020 (Torres).



\end{document}